**Nature Sustainability Comment**

**PRE-PRINT DRAFT – MAY DIFFER SLIGHTLY FROM FINAL PUBLICATION**

**Title: Sustainable Skies and the Earth-Space Environment**

Authors: A. Williams[1], A. Boley[2], G. Rotola[3], R. Green[4]

[1]International Astronomical Union, Centre for the Protection of the Dark and Quiet Skies from Satellite Constellation Interference / European Southern Observatory. Andrew.Williams@eso.org
[2]Department of Physics and Astronomy, University of British Columbia, 6224 Agricultural Rd., Vancouver BC, V6T 1Z1, Canada. aaron.boley@ubc.ca
[3] Sant'Anna School of Advanced Studies, Piazza Martiri della Libertà, 33, 56127 Pisa PI, Italia. giuliana.rotola@santannapisa.it
[4]International Astronomical Union, Centre for the Protection of the Dark and Quiet Skies from Satellite Constellation Interference / University of Arizona, Steward Observatory

**The rapid launch of hundreds of thousands of satellites into Low Earth Orbit will significantly alter our view of the sky and raise concerns about the sustainability of Earth's orbital space. A new framework for sustainable space development must balance technological advancement, protection of space environments, and our capacity to explore the Universe.**

From the dawn of human history, we have been guided by the heavens. Our ancestors experienced thousands of millennia of pristine night skies, which played a tremendous role in shaping human society. Our religions, cultures, timekeeping, and celebrations are all shaped by the motions of celestial objects. But now, our window to the Universe is changing.

Until very recently, the count of satellites in low-Earth orbit (LEO) was in the low thousands. If the trend of satellite deployment persists over the next ten years, however, the number of LEO satellites is on track to exceed 100,000. At any given location and moment on Earth, several thousand satellites will be detectable above the horizon. Along with debris and abandoned orbital infrastructure, the sky could come alive with streaks and flashes, masking and confusing systematic searches for astronomical sources. And for a radio telescope, the sky would be full of noise and artificial signals.

The impacts on astronomy are but a symptom of the rapid growth in space traffic, which while realizing benefits brings a set of startling negative externalities. The present trajectory of satellite deployment presents a formidable sustainability challenge and the global space governance system currently falls short in its management of this challenge. It is time to reconsider the principles governing space sustainability.

Our space governance system is underpinned by the Treaty on Principles Governing the Activities of States in the Exploration and Use of Outer Space, including the Moon and Other Celestial Bodies, otherwise known as the Outer Space Treaty (OST). The OST, which entered into force in 1967 and now counts over 100 signatories, was drafted in an era when the scale and complexity of space endeavours bore little resemblance to today's space economy, characterized by thousands of satellites, commercial ventures, and ambitious projects like space-based solar power, asteroid mining, and in-space manufacturing.

The OST grants broad freedoms to States for outer space exploration and scientific investigation, coupled with restrictions prohibiting nuclear weapons in space, military bases on celestial bodies, and national appropriation of outer space. While the treaty emphasizes benefits for all humanity and consideration of due regard for other States' interests, States tend to take a view that actions are unrestricted in all but the most severe violations of provisions within the corpus of space law. This is fundamentally at odds with sustainable development, which requires not only exercising restraint but also recognizing the existence of limits.

While governments and operators have long been aware of the potential dangers of growing space traffic and debris, the current circumstances reveal that a de facto unrestricted development has indeed transpired. Current estimates put the number of trackable debris objects in space at over 34,500, but the technology and regulation to remove or deorbit space debris remain nascent. Moreover, the number of lethal but nontrackable objects is over one million[1]. Collisions between satellites or with pieces of space junk increase the potential for knock-on collisions, generating additional space debris and making it more difficult and dangerous to operate in LEO. Defunct satellites and rocket parts contribute substantially to this risk, being hulking pieces of unmanoeuvrable mass. The growing debris and satellite congestion threaten human activity in space, including the International Space Station and Chinese Tiangong, which need to manoeuvre to avoid collisions. The many emerging commercial projects in LEO will have to contend with this congestion, and as operators fill up certain orbital regions with tens of thousands of objects, this can effectively deny the use of these orbits for other users or make operations in those regions risky. The uncontrolled re-entry of thousands of satellites into Earth's atmosphere could also pose a hazard to people and property on the ground and to airplanes in flight[2]. Re-entries could also alter Earth's upper atmosphere chemistry in non-trivial ways, owing to ablation products that are wildly different from the background flux of meteoroids.[3] [4] [5]

The space community has not been idle in looking for solutions. In 2002, several national space agencies and the European Space Agency, under the framework of the Inter-Agency Space Debris Coordination Committee (IADC), published the first set of voluntary best practices to minimise space debris[6]. Industry-led initiatives have proposed best practice guidelines to operate spacecraft safely, coordinate among operators, and reduce space debris generation.[7] A coalition of universities with the involvement of the European Space Agency and the World Economic Forum initiated the Space Sustainability Rating project to rate operators according to transparent, data-based assessments of the level of sustainability of space missions and operations.[8]

Regulations are also changing. For example, the US Federal Communications Commission, which acts as a de facto US mission authoriser, shortened the required deorbit time for defunct satellites from 25 to 5 years.

*Figure 1: Projections of space objects onto Earth. Top. Current distribution of active and defunct satellites and abandoned rocket bodies, according to the US Space Command (accessed: 27 August 2023). Bottom. Approximately 65,000 filed satellites (but subject to change) by SpaceX, OneWeb, GuoWang, and Amazon. Many other satellites are being filed. Only proposed LEO satellites are shown in the lower panel. Outlines do not imply the expression of any opinion concerning the legal status of any country, territory, city, or area or of its authorities, or concerning the delimitation of its frontiers or boundaries.*

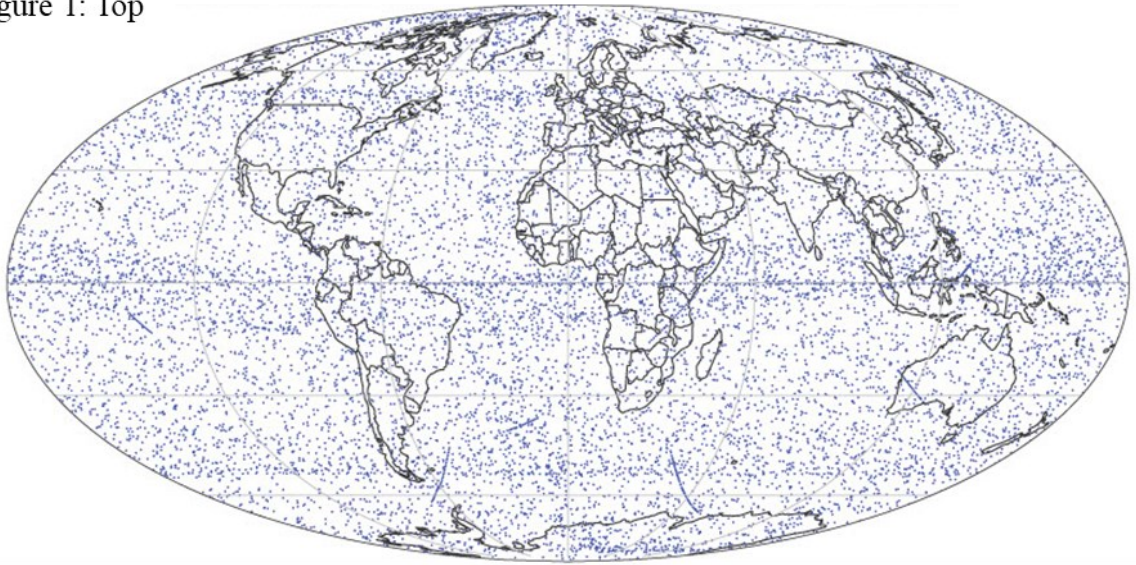

Figure 1: Top

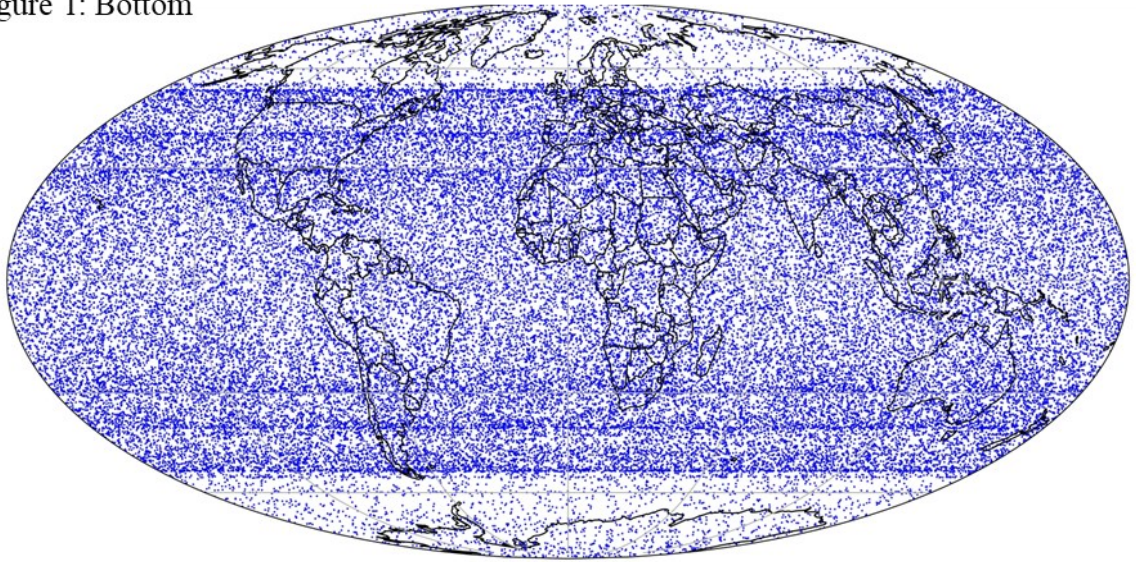

Figure 1: Bottom

The astronomy community has taken note of these activities and has sought to raise the profile of "Dark and Quiet Skies" as a component of space sustainability.[9] The strategy first recognises the geopolitical interests at stake in LEO and that satellite operators have a legal right to operate under international law. The astronomy community is thus collaborating with industry to identify mitigations. This work has resulted in experimental design modifications by a few companies, including SpaceX and Amazon Kuiper[10]. Such efforts have made progress toward the brightness guidelines established by the International Astronomical Union (IAU), which include setting the brightness limit below naked eye visibility[11]. Nonetheless, all satellites remain fully detectable by large aperture telescopes.

Second, the astronomy community is taking steps to raise awareness in policymaking circles. The IAU established a Centre for the Protection of Dark and Quiet Skies from Satellite Constellation Interference. The IAU, along with other astronomy organisations, have successfully introduced the Dark and Quiet Skies topic on the agenda at the UN, G7 and EU Council. The final and longer-term step is to turn this industry practice and political capital into regulation that seeks to address the major impacts on astronomy, and more generally, observing the sky.

This story may seem familiar, having previously unfolded in industries like oil, plastics, chemical fertilisers, insecticides, and CFCs. The pattern typically involves rapid technological development accompanied by innovative applications that bring societal benefits. Eventually, adverse repercussions and negative impacts emerge, including environmental pollution threats to other industries, ecological harm, and potential harm to human well-being. Awareness of these downsides often lags technological progress. Recognising the problem, industry may respond with voluntary measures and sustainability initiatives. Environmental groups mobilise to raise political awareness, while industry continues its activities largely unabated. Given the relatively nascent state of industrialisation in Earth's orbital realm, we have a window of opportunity to avoid mistakes of the past. We can develop a space economy in a way that safeguards its future and preemptively addresses unintended consequences, such as preserving the unspoiled night sky.

The United Nations Committee on the Peaceful Uses of Outer Space (COPUOS) is a forum of 102 States with the mandate to foster international cooperation arising from the exploration of outer space. In 2019, following a decade of negotiations within a COPUOS working group, member States reached consensus on the Guidelines for the Long-term Sustainability of Outer Space Activities (LTS Guidelines). The document, which covers a wide range of topics, such as policy, safety, cooperation, and scientific research, includes 21 guidelines that aim to foster sustainable practices for conducting outer space activities. While the LTS Guidelines are not legally binding, they do draw upon foundational principles in international law, including space law. Importantly, states are studying how to implement them in their national legislation and standards.[12]

While the LTS Guidelines are a notable achievement in diplomacy and an important norm-setting mechanism for future space policies, we argue their foundational assumption – that a domain can be developed perpetually – is inherently flawed. Indeed, the LTS Guidelines define sustainability as the ability to maintain the conduct of space activities indefinitely into the future in a way that grants equitable access to benefits and preserves the space environment for future generations. The latter is laudable but will be undercut when the former fails.

There are thresholds beyond which further development becomes unsustainable, leading to degradation and adverse impacts. For example, a carrying capacity can be defined for any orbit, in principle, taking into account the density of objects, technological capabilities, and collision risk levels[13]. Another example is the background electronic noise produced by satellites, for which a threshold exists beyond which observations of faint cosmic sources are impossible. This underscores the necessity of re-evaluating the meaning of sustainable development in the space sector. Proponents of traditional sustainability may suggest reducing development rates, but this could merely postpone reaching capacity limits. A more profound shift in our relationship with development is required.

The current corpus of voluntary guidelines and industry charters strongly assume that we can "tech" our way towards long-term sustainability. Merely relying on technologies such as AI-based collision avoidance for satellites or on-orbit space debris removal, without considering its potential consequences, is insufficient. Technological progress alone may exacerbate issues, as seen with increasingly dense orbital traffic and debris.

To deal with emergent harms such as the impact on astronomy, we advocate for a new framework for the development of space – one that builds on the UN ideas of sustainable development, but also recognises that, as a resource is consumed, the rate of development must be reduced or fundamentally changed. An equilibrium state, analogous to ecological sustainability, should be the goal. This requires reframing space governance in a way that is both adaptive and attuned to the space environmental system.

Adaptive governance of ecosystems involves a flexible approach to governance that swiftly responds to environmental changes, incorporating new information and adapting to evolving conditions. Rooted in ecological principles, this approach recognizes the interconnectedness of all ecosystem components, prioritizing the overall balance, diversity, and resilience. In other words, solutions for space traffic management and debris remediation must recognise their broader impacts on the Earth's atmosphere, astronomical observations, and the ability of other uses to access space.

Contrasting current sustainability approaches focusing on specific technological solutions, an adaptive approach for space sustainability demands shared responsibility, stakeholder collaboration, engaging diverse parties in decision-making processes, fostering iterative learning, and generating varied solutions for different scenarios. This requires regulatory and policy measures and incentives at national and international levels, coupled with bold leadership and international diplomacy. Space policymakers should pay heed to the numerous treaties addressing pollutants or conservation and the lessons learnt in ecosystem management. Drawing from these extant approaches, a non-exhaustive list of measures can be applied to the space domain as illustrated in Figure 2.

As humanity's space capabilities expand, so do the challenges. The proliferation of satellite constellations exemplifies the tension between innovation and consequences, demanding a space policymaking and governance system adaptable to evolving space activities and their unforeseen impacts. To ensure a balance between progress and preservation, space sustainability strategies must draw insights from the corpus of knowledge in environmental sustainability to secure both technological development and the longevity of space "ecosystems". Retaining an unrestricted view of our window to the Universe and the pursuit of knowledge in astronomy hinges on striking this equilibrium.

*Figure 2: An Adaptive Governance-informed plan for Space Sustainability*

> 1) Foster transparent and current awareness of space object population, sourced from commercial and government entities, to enhance public access to space environment information;
> 2) Use a global forum to coordinate knowledge about space objects and operations across institutional levels, with existing dialogues between operators and governments serving as a foundation;
> 3) Ensure that every regulatory authorisation of new space objects is informed by a long-term extrapolation of aggregate contribution to space debris risk, light pollution, and deposition of metals in the Earth's upper atmosphere;
> 4) Prioritize research on environmental thresholds and limits to establish sustainability targets and effective monitoring mechanisms;
> 5) Implement measures to protect vulnerable space regions or establish restricted zones, moving away from unrestricted access and utilization principles;
> 6) Design transparent and equitable financial incentives to enhance sustainability technologies and fund collaborative clean-up initiatives.

**Author Contribution Statement**: AW led the author team, conceived of the article framework, drafted parts of the manuscript and created Figure 2. AB drafted parts of the manuscript and conducted the simulations to produce Figure 1. GR and RG drafted parts of the manuscript.

---